\documentclass[submission,creativecommons,noncommercial,sharealike]{eptcs}
\providecommand{\event}{F-IDE 2018}

\usepackage{breakurl}
\usepackage[strings]{underscore}
\usepackage{listings}
\usepackage{fancyvrb}
\usepackage{graphicx}
\usepackage{float}
\usepackage{caption}

\title{A Notebook Format for the Holistic Design of Embedded Systems (Tool Paper)}

\author{
    Spencer Park
    \institute{
        McMaster University\\ 
        Hamilton, Canada
    }
    \email{parksj6@mcmaster.ca}
\and
    Emil Sekerinski
    \institute{
        McMaster University\\ 
        Hamilton, Canada
    }
    \email{emil@mcmaster.ca}
}

\def\titlerunning{Notebooks for Embedded Systems}
\def\authorrunning{Spencer Park and Emil Sekerinski}

\providecommand{\tightlist}{%
      \setlength{\itemsep}{0pt}\setlength{\parskip}{0pt}}

\hypersetup{
      colorlinks=true,
      linkcolor=blue,
      urlcolor=blue,
      citecolor=blue,
      anchorcolor=blue,
}

\begin{document}
\maketitle

\begin{abstract}
    This paper proposes the use of notebooks for the design documentation
and tool interaction in the rigorous design of embedded systems.
Conventionally, a notebook is a sequence of cells alternating between
(textual) code and prose to form a document that is meant to be read
from top to bottom, in the spirit of literate programming. We extend the
use of notebooks to embedded systems specified by pCharts. The charts
are visually edited in cells inline. Other cells can contain statements
that generate code and analyze the charts qualitatively and
quantitatively; in addition, notebook cells can contain other
instructions to build the product from the generated code. This allows a
notebook to be replayed to re-analyze the design and re-build the
product, like a script, but also allows the notebook to be used for
presentations, as for this paper, and for the inspection of the design.
The interaction with the notebook is done through a web browser that
connects to a local or remote server, thus allowing a computationally
intensive analysis to run remotely if needed. The pState notebooks are
implemented as an extension to Jupyter. The underlying software
architecture is described and the issue of proper placement of
transition labels in charts embedded in notebooks is discussed.

\end{abstract}

    \hypertarget{introduction}{%
\section{Introduction}\label{introduction}}

This work addresses the design documentation and user interface aspects
in the rigorous design of embedded systems.

In the 80's, Knuth argued that programs should be written as if they are
the work of literature: ``Let us change our traditional attitude to the
construction of programs: Instead of imagining that our main task is to
instruct a computer what to do, let us concentrate rather on explaining
to human beings what we want a computer to do''
~\cite{Knuth84LiterateProgramming}. In his system for \emph{literate
programming}, there is a single source file containing both the
executable code and explaining prose. One tool, \emph{tangle}, extracts
the code for submission to a compiler and another tool, \emph{weave},
processes the embedded markup instructions to generate a hyperlinked,
pretty-printed document using TeX. Literate programming has been used
for compilers~\cite{FraserHanson95RetargetableCCompiler}, scientific
software~\cite{Nedialkov06VNODELP}, \href{http://ulixos.org/}{operating
system}~\cite{link:ulixos}, cryptography~\cite{Klein13StreamCiphers} as
well as Knuth's TeX system~\cite{Knuth86TypesettingB}. The ideas have
also influenced
\href{http://spivey.oriel.ox.ac.uk/corner/Fuzz\%20http://spivey.oriel.ox.ac.uk/mike/fuzz/}{formal
specification languages}~\cite{link:fuzz} and
\href{https://www.haskell.org/onlinereport/literate.html}{functional
programming languages}~\cite{link:literate-haskell}. The documentation
facilities of languages like
\href{http://www.oracle.com/technetwork/java/javase/documentation/index-137868.html}{Java}
~\cite{link:javadoc} also draw from literate programming, even if there
the prose is embedded in the code rather than on equal footing. The
notebooks of
\href{http://reference.wolfram.com/language/\#NotebookDocumentsAndPresentation}{Mathematica}
~\cite{link:mathematica} and \href{http://jupyter.org/}{Jupyter}
~\cite{link:jupyter} and the worksheets of
\href{https://www.maplesoft.com/support/help/Maple/view.aspx?path=worksheet}{Maple}
~\cite{link:maple} add \emph{interactivity} to literate programming: code
fragments can be executed right in the editor and their results, which
can be text, formulae, or diagrams, are displayed inside the document.
In Jupyter, prose can be formatted with \emph{markdown}, an HTML-based
markup language, and is pretty-printed interactively, thus eliminating
the edit-generate-process cycle of Knuth's \emph{weave}.

System models are meant to be critiqued by humans as well as
mechanically processed. It is therefore fitting to consider a literate
approach to designing those. This paper reports on the design decisions
of the re-implementation of the \emph{pState} user interface with
notebooks~\cite{NokovicSekerinski13pState}. The pState tool supports the
\emph{holistic design} of embedded systems with \emph{pCharts}, a visual
formalism for embedded systems:

\begin{itemize}
\item
  The state of the system is described by hierarchical and concurrent
  states together with variables within those states; transitions
  between states can be triggered by (external and internal) events, can
  be timed (deterministic, nondeterministic within an interval,
  uniformly distributed within an interval, exponentially distributed),
  and can be probabilistic. For (sub-) charts without probabilistic
  transitions, executable code can be generated. Thus pCharts can model
  both the system under development as well as its environment
~\cite{NokovicSekerinski14TimedpCharts}.
\item
  Charts can be analyzed qualitatively and quantitatively. For the
  qualitative analysis, \emph{invariants} can be attached to
  hierarchical states and the correctness of the transitions with
  respect to the \emph{accumulated invariant}
~\cite{Sekerinski09StateInvariants} is verified.
\item
  For the quantitative analysis, \emph{costs} can be attached to
  transitions and states (state costs accumulate linearly with time).
  Queries for the (minimal and maximal) reachability probability and for
  expected (maximal and minimal) costs can be attached to states
~\cite{NokovicSekerinski15HolisticEmbeddedSystem}.
\item
  Maximal execution times of transitions can be specified and are
  checked against the generated code, taking the scheduler into account
~\cite{NokovicSekerinski15WCETinvariants}.
\item
  Comments can be embedded in the charts and are preserved in the
  generated executable code. If they are attached to states and
  transitions, then they are placed in the generated code where the
  corresponding variables are declared or transition is taken
~\cite{NokovicSekerinski15HolisticEmbeddedSystem}.
\end{itemize}

Internally, pState first transforms the pChart model into
\emph{probabilistic guarded commands with priority}. That representation
serves for further transformations:

\begin{itemize}
\item
  For (sub-) charts without probability, C code and Arduino code
  (Arduino is experimental) can be generated
~\cite{NokovicSekerinski17Tags}.
\item
  For (sub-) charts without probability, PIC assembly code for ATMega
  microcontrollers can be generated and the worst-case execution time of
  the generated code can be analyzed~\cite{NokovicSekerinski17Tags}.
\item
  Whole charts are translated into the guarded command language of the PRISM
 probabilistic model checker~\cite{KwiatkowskaNormanParker11PRISM4}.
  Invariants and quantitative queries are translated to PRISM PCTL
  formulae.
\item
  Transition guards are compiled into formulae for the Yices SMT solver
~\cite{Dutertre14Yices} to detect infeasible paths; this is used to
  improve the WCET analysis of the generated code.
\end{itemize}

Interaction with notebooks can take place in different ways:

\begin{itemize}
\item
  Notebooks can be edited and executed through a web browser that
  connects to a (local or remote) Jupyter server running a Python kernel
  with pState.
\item
  Notebooks can be rendered through GitHub or a service like
  http://nbviewer.jupyter.org; editing and execution is not possible,
  but all parts of the notebook can be inspected. For example, this
  notebook can be viewed at its
  \href{https://gitlab.cas.mcmaster.ca/lime/pstate-jupyter/blob/master/docs/submission.ipynb}{GitLab
  Repository}~\cite{link:pstate-jupyter-sub-notebook}. At the time of
  writing, the GitLab notebook renderer has several limitations. One is
  that only png (bitmap) images work as expected, which necessitates
  that all images in this notebook are bitmaps. A more fundamental
  limitation is that citations and references to figures are not
  displayed properly in notebooks, but are in the generated pdf file.
\item
  Notebooks can be converted to pdf; editing and execution is not
  possible and hidden cells will be suppressed. A pdf version of this
  notebook (possibly the version the reader is perusing) has been
  converted with a dedicated
  \href{https://gitlab.cas.mcmaster.ca/parksj6/pstate-jupyter-eptcs-exporter}{converter}
~\cite{link:pstate-jupyter-exporter} for the purpose of publication.
\end{itemize}

For example, a new chart is created by calling
\texttt{pChart("filename")} in a code cell and assigning it to a Python
variable. The JSON representation of the chart would then be stored in the file named 
\texttt{\_\_pcharts\_\_/filename.pchart}. Below, a new chart is assigned
to \texttt{sender\_receiver}. Displaying the value of
\texttt{sender\_receiver} opens then a cell for editing; the menu with
the commands for drawing AND, XOR states, transitions, etc. appears only
interactively and is not visible in this notebook.

\begin{lstlisting}[language=python]
from pstate import pChart
sender_receiver = pChart('sender_receiver')
generated_intermediate_code = sender_receiver.code()
sender_receiver # display the editor

\end{lstlisting}

\begin{figure}
\centering
\includegraphics[scale=0.75]{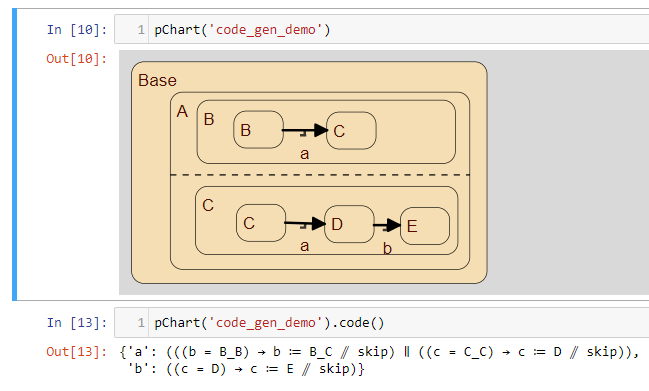}
\caption{Intermediate event code generation inside a Jupyter notebook
\label{fig:code-gen-demo}}
\end{figure}

    The chart is a Python object on which methods can be called: \texttt{sender\_receiver.code()}, for example, would generate intermediate code for
the chart. Other methods can then inspect invariants, events, and costs.
Figure \ref{fig:code-gen-demo} is a screenshot where the chart
\texttt{code\_gen\_demo} is displayed and in the following cell, the
intermediate code is generated and displayed.

    \hypertarget{the-pstate-jupyter-architecture}{%
\section{The pState-Jupyter
Architecture}\label{the-pstate-jupyter-architecture}}

\begin{figure}
\centering
\includegraphics{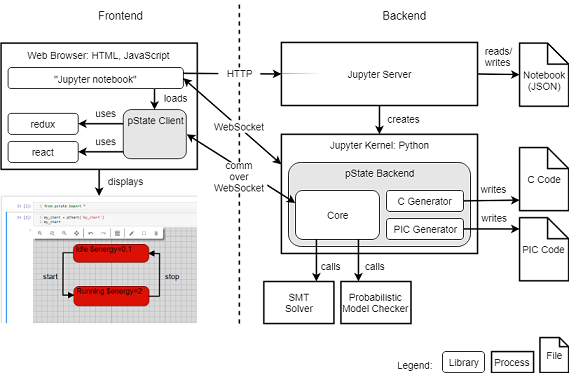}
\caption{Notebook architecture \label{fig:notebook-architecture}}
\end{figure}

Jupyter notebooks are a document format that interleave prose in
markdown cells, code cells, and code execution results. Notebooks,
including the code execution results, are stored in JSON files. The
three major components that make up the interactive environment for
running notebooks are the \emph{server}, the \emph{frontend}, and the
\emph{kernel}. An architecture diagram is depicted in Figure
\ref{fig:notebook-architecture}.

The server interacts with the file system to store notebooks.
Communication between the frontend and the server is done via HTTP
requests/responses. Frontends may also request the creation of a kernel
and the server has the responsibility of spawning the appropriate kernel
process and providing it with connection details so that the kernel and
frontend can connect directly. Kernels exist for numerous programming
languages. For pState, we use the IPython kernel.

The frontend provides the graphical user interface for viewing and
editing the notebook prose and code. When connected to a kernel it may
also send a code cell to the kernel for execution; for pState, these are
Python commands that are interpreted by the Python kernel. The result is
sent to the frontend and displayed. Frontends commonly support rendering
image formats such as PNG, JPEG, and SVG, as well as richer formats
including LaTeX equations, HTML, and JavaScript.

The kernel is a process responsible for executing notebook code, as well
as providing other language features such as code completion. The
communication with the frontend is via the Jupyter messaging protocol.
This communication is two-way, allowing the frontend to make requests to
the kernel and vice-versa.

Embedding pState into a Jupyter notebook requires the coordination of
two components, one on the frontend for viewing and graphically editing
pCharts and another on the backend for compiling, analyzing, and
generating code for pCharts.

The frontend is a web application written in
\href{https://www.typescriptlang.org/}{TypeScript}
~\cite{link:typescript}, which is compiled to JavaScript to run in a
browser, and built on the \href{https://reactjs.org/}{React}
~\cite{link:react} and \href{https://redux.js.org/}{Redux}
~\cite{link:redux} frameworks. In a Redux application, the entire state
of the application is a single JavaScript object. This state is
immutable and modified by pure \emph{reducers} that are triggered via
dispatched \emph{actions}. Additionally the state and actions are JSON
serializable, making synchronization and remote action dispatching
possible. When the state changes, React renders a virtual view of the
changed state, compares it to the concrete view, and makes the necessary
changes to the concrete view so that it matches the virtual view. The
concrete view is an HTML element with inline SVG for most of the chart
drawing.

The backend component is a Python library that lives in the kernel, as
the pState interface must be accessible to user code written in the
notebook. Code written in the notebook is sent to the Python kernel for
execution. Every pChart is a Python object; ``displaying'' the object
opens the inline editor. Methods of a pChart object are called for
analysis and code generation.

The frontend is connected to the backend via the Jupyter messaging
protocol's \texttt{comm} messages. As such, the components communicate
through asynchronous message passing. The backend sends chart updates,
dispatches user interface actions, and sends chart errors or analysis
results to the frontend to display to the user. The frontend sends
compilation or analysis requests as well as chart modifications to the
backend.

    \hypertarget{interface}{%
\section{Interface}\label{interface}}

In the notebook context the graphical editor is to be used when it is
helpful. All of the functionality of the editor is also accessible via
the Python object residing in the kernel. This allows for scripting
changes to the diagram in addition to graphically editing it, which in
the notebook format is side by side.

Future work on the backend will widen the pChart API to support
programatically varying constants. With this feature we anticipate
Jupyter users will bring the libraries and tools they are familiar with
to integrate with pState. Such examples include
\href{https://github.com/matplotlib/jupyter-matplotlib}{Matplotlib}
~\cite{link:matplotlib} or
\href{https://github.com/bloomberg/bqplot}{bqplot}~\cite{link:bqplot}
for plotting analysis results and
\href{https://github.com/jupyter-widgets/ipywidgets}{ipywidgets}
~\cite{link:ipywidgets} to create interactive interfaces for modulating
constants defined within their pCharts.

\begin{figure}
\centering
\includegraphics[scale=0.53]{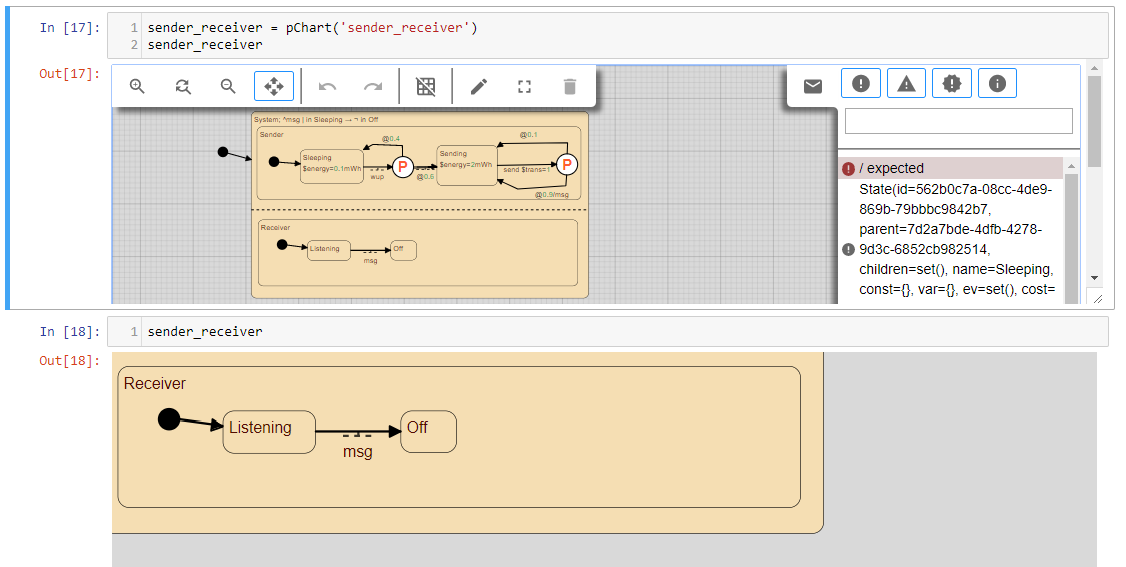}
\caption{Duplicate editors open for a single pChart
\label{fig:sender-receiver:multiple-views}}
\end{figure}

The notebooks format necessitates that charts are as wide as all other
cells, which are typically of the width of a printed page, rather than
filling the whole screen. The editor was designed to not overflow the
space it is allocated to render in. It is fully contained and therefore
many editors may be present within the same webpage. Additionally any
menus are kept to a minimum and hidden when not being edited to leave as
much space for the chart. To allow for large charts, a single chart (or
rather the underlying Python object) can have multiple views, which are
kept automatically synchronized. These views can provide an overview or
``zoom'' in particular states of the chart. Even then, the layout of
charts is critical for readability. Figure
\ref{fig:sender-receiver:multiple-views} is a screenshot of a chart that
is displayed twice with the user working in the top editor causing it to
be in focus making the menus visible.

In the spirit of using the visual editor as a tool it implements some
features for making editing quicker, namely automatically positioning
labels and drawing concurrent state separating lines with plans for
extending this list in the future with automatic reformating and
connection path routing.

\begin{figure}
\centering
\includegraphics[scale=0.5]{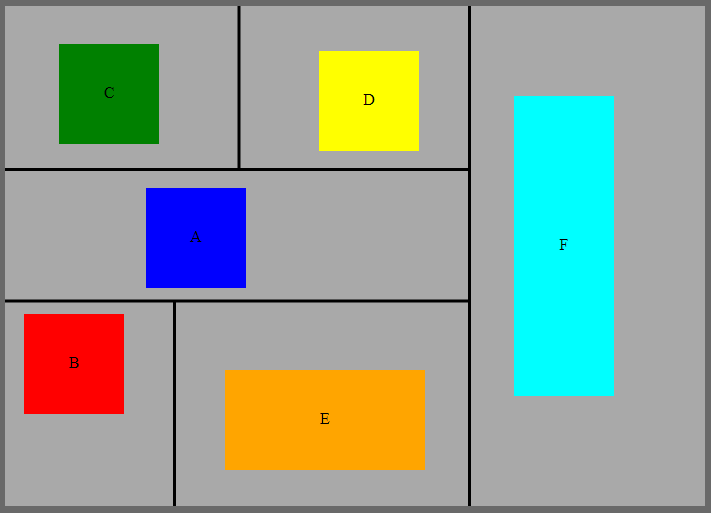}
\caption{Automatic ``concurrent state'' child splitting algorithm
visualizaion \label{fig:box-division:split-all}}
\end{figure}

Concurrent states (AND states) visually separate their children with a
dashed line. These lines are drawn automatically by recursively
splitting groups of child states with a dashed line spanning the entire
group across a single axis. Any groups left are split across the
alternate axis until a group cannot be split further. Figure
\ref{fig:box-division:split-all} shows how the algorithm splits the
child boxes A-F.

Transitions between states are broken down in \emph{connections} between
states and probabilistic/conditional pseudo-states. These serve for
splitting a transition depending on a probability or on a condition.
Like commonly done in drawing editors, the frontend requires states and
pseudo-states to be positioned manually, but automatically places the
labels of connections. The frontend implements a variation of an
algorithm by Kakoulis and Tollis for labeling edges in hierarchical
drawings~\cite{KakoulisTollis1997LabelingEdges}. It consists of three
steps,

\begin{enumerate}
\def\labelenumi{\arabic{enumi}.}
\tightlist
\item
  finding viable label positions,
\item
  rating the positions on how good their placement is, and
\item
  choosing a subset of those positions (one for each label) that reduces
  the total ambiguity in the relationship of label to connection.
\end{enumerate}

Positions that collide with objects in the chart, including states and
connections, are removed from the set of potential positions. As pCharts
are \emph{hierarchical} state charts, nested states will have label
positions that overlap their parent states. When removing labels
that collide with objects in the chart, the algorithm must not consider
collisions with parent objects that are conceptually \emph{behind} the
label. The algorithm bases this decision on where the source state of
the connection exists in the state hierarchy.

\begin{figure}
\centering
\includegraphics{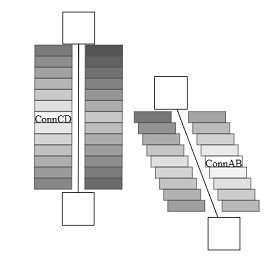}
\caption{Edge label placement computation visualization
\label{fig:elp:force-middle-edge}}
\end{figure}

The costs is based on a force directed calculation for pulling towards
the centroid of a connection path and pushing away from other labels.
This algorithm may result in placements that are still of high cost. In
these cases the editor renders a linking dotted line from the label to
it's connection to help reduce the ambiguity in the resulting placement.
In any case the user is also free to manually position any labels that
are not placed to their standards. Figure
\ref{fig:elp:force-middle-edge} shows the potential label positions
along with their computed costs (darker implies a ``worse'' placement)
and resulting placement decision.

    \hypertarget{conclusions}{%
\section{Conclusions}\label{conclusions}}

This project was initiated to explore if it is possible to adopt a
notebook format for embedded system design with hierarchical state
charts. Using Jupyter notebooks, this turns out to be feasible, even if
with considerable complexity: a separation of the frontend and backend
is necessary and the frontend alone consists of 10,000 lines of
TypeScript and uses two JavaScript frameworks. The advantages of the
separation are that the backend can run on a possibly remote server and
only a browser is needed for connecting to the backend; from the user's
point, the complexity is not visible but the benefits are obvious. The
frontend implementation took approximately seven months which includes
roughly two months spent learning about web technologies. Connecting the frontend
to the backend was quite painless due to the design decisions made when
implementing the editor, mainly using Redux's design patterns including
immutable state and rendering the UI based on a single state object,
allowing the efficient replacement of the entire state for
synchronization. As such the complexity for a reader to do the same is
very much dependent on the implementation of the web interface.

At the time of writing, the backend still misses the integration of
PRISM, Yices, and most of the code generators of the original pState
implementation; we hope to report on using the notebooks format for full
development soon. Additionally there are plans to integrate a
simulation/animation tool into the frontend as well as bringing the
editor into the newer interface
\href{https://github.com/jupyterlab/jupyterlab}{JupyterLab}
~\cite{link:jupyterlab} which was recently (late February 2018) released
as a stable beta.

Notebooks have been proposed for documenting \emph{reproducible
research}~\cite{KluyverEtAl16Jupyter}. The authors give numerous
examples, but also write ``It is not yet very practical to write
academic papers themselves as notebooks, but we are working towards
this''. We have managed to generate a pdf version of this paper from a
notebook.

More importantly, we believe that notebooks can support a rigorous
design process: the notebook format serves uniformly for designing,
documenting, and for building the executable code. As they are
self-contained and can be replayed, they can be inspected by third
parties. Notebooks may therefore prove to be suitable for certification.

\nocite{*}
\bibliographystyle{eptcs}
\bibliography{NotebooksForEmbeddedSystems}

\end{document}